\documentclass[pageno]{jpaper}

\usepackage[normalem]{ulem}
\usepackage{amsmath,amssymb,amsfonts}
\usepackage{algorithmic}
\usepackage{textcomp}
\usepackage{xcolor}
\usepackage{subfiles}
\usepackage{multicol}
\usepackage{caption}
\usepackage{authblk}
\usepackage[flushleft]{threeparttable}
\usepackage{graphicx}

\usepackage[bottom]{footmisc}
\def\showcorrections{1}
\ifx\showcorrections\undefined
    \newcommand{\cdel}[1]{}
    \newcommand{\cdeleq}[1]{}
    
\else
    \newcommand{\cdel}[1]{\textcolor{red}{#1}}
    \newcommand{\cdeleq}[1]{\textcolor{red}{#1}}
    
\fi

\author[1]{Konstantinos Iordanou}
\author[2]{Timothy Atkinson}
\author[3]{Emre Ozer}
\author[3]{Jedrzej Kufel}
\author[3]{John Biggs}
\author[1]{Gavin Brown}
\author[1]{Mikel Luj\'an}
\affil[1]{Department of Computer Science, University of Manchester, UK \authorcr\textit{\{firstname.lastname\}@manchester.ac.uk}}
\affil[2]{NNAISENSE, Switzerland \authorcr\textit{\{firstname.lastname\}@nnaisense.com}}
\affil[3]{Pragmatic Semiconductor, Cambridge, UK \authorcr\textit{\{eozer, jkufel, jbiggs\}@pragmaticsemi.com}}

\begin{document}

\title{Tiny Classifier Circuits: \\Evolving Accelerators for Tabular Data}

\date{}
\maketitle

\thispagestyle{empty}

\begin{abstract}

A typical machine learning (ML) development cycle for edge computing is to maximise the performance during model training and then minimise the memory/area footprint of the trained model for deployment on edge devices targeting CPUs, GPUs, microcontrollers, or custom hardware accelerators. 

This paper proposes a methodology for automatically generating predictor circuits for classification of tabular data with comparable prediction performance to conventional ML techniques while using substantially fewer hardware resources and power. The proposed methodology uses an evolutionary algorithm to search over the space of logic gates and automatically generates a classifier circuit with maximised training prediction accuracy. Classifier circuits are so tiny (i.e., consisting of no more than 300 logic gates) that they are called ``Tiny Classifier'' circuits, and can efficiently be implemented in ASIC or on an FPGA. 

We empirically evaluate the automatic Tiny Classifier circuit generation methodology or ``Auto Tiny Classifiers'' on a wide range of tabular datasets, and compare it against conventional ML techniques such as Amazon's AutoGluon, Google's TabNet and a neural search over Multi-Layer Perceptrons. Despite Tiny Classifiers being constrained to a few hundred logic gates, we observe no statistically significant difference in prediction performance in comparison to the best-performing ML baseline. When synthesised as a Silicon chip, Tiny Classifiers use 8-18x less area and 4-8x less power. When implemented as an ultra-low cost chip on a flexible substrate (i.e., FlexIC), they occupy 10-75x less area and consume 13-75x less power compared to the most hardware-efficient ML baseline. On an FPGA, Tiny Classifiers consume 3-11x fewer resources.

\end{abstract}


\section{Introduction}
The relentless successes of Deep Neural Networks (DNNs), in achieving near (or better than) human accuracy for important application domains has created tremendous research and industrial momentum. Although originally much success was based on Convolutional Neural Networks and harnessing the availability of large labelled datasets of images, the successes have expanded to various other tasks and associated neural architectures (e.g., recurrent and transformers for Natural Language Processing). These large datasets are mainly images, audio or text. This kind of data can be characterised as homogeneous data.

Given the momentum gathered and the existence of common computational kernels across the different kinds of DNNs, we are witnessing a myriad of hardware accelerators for \textit{inference} as well as \textit{training} of DNNs. In both scenarios, the most common approach for these accelerators is to be programmable hardware with specialized datatypes and computations, rather than being a task-specific circuit. As DNNs have evolved, their computation has evolved from dense tensor operations towards increased sparsity.

To sum up, the current status quo separates the development of the specific DNN for a particular task from the process of developing the hardware accelerator for the training, or inference, of the specific DNN. In more general terms, the current best practice considers a Machine Learning (ML) technique which generates a model, where the training and the execution of the model versus the design and optimization of the hardware accelerator are isolated; at best a co-design happens. Nonetheless, both development activities intrinsically involve optimization processes. Thus, a reasonable question to postulate would be: \textit{Could we develop a supervised learning technique that takes tabular data as input, and generates a circuit representation for classification behaving like an ML model?}

Our main contribution is to address this question by presenting a methodology to automatically generate classification circuits directly from tabular data. In contrast to homogeneous data (image, text), we focus on tabular data which, for example, can combine numerical and categorical data (heterogeneous). DNNs excel at capturing the spatial or semantic relationship in images or speech data. However, for tabular data, the correlation among the features is weaker, and the features have no intrinsic positional information. Hence, tabular data is an active research area for DNNs \cite{tabnet, node, tabular_data, regularization_is_all_you_need}.

Such heterogeneous data are ubiquitous \cite{tabular_data}, with many associated real-world applications. The paper describing Google's TabNet \cite{tabnet} refers to tabular data as ``the most common type of data in real world AI''. Important use-cases for tabular data appear, for example, in healthcare \cite{10.1145/3285029}, where various numerical and categorical descriptors of patients can be used to infer suggestions for medication \cite{zhang2015cadre,bao2016intelligent} and personalised treatments. Of particular relevance, tabular data often exists in resource-limited scenarios suited to low-power ML also known as tinyML \cite{lee2016integrating,armpit, li2018implemented, kumar2018novel}.

The fundamental reason behind the benefits of our methodology is two-fold. Firstly, our boolean function representation is otherwise known as a ``decision tree'', in ML, and thus inherits favourable properties of this representation. A series of recent studies has indicated that decision trees outperform Deep Learning on tabular data, notably Grinsztajn et al. \cite{grinsztajn_et_al} observe that tree-based models have a natural advantage in such data. Specifically, they explain:

\begin{quote}
    {\em ``This superiority is explained by specific features of tabular data: irregular patterns in the target function, uninformative features, and non rotationally-invariant data where linear combinations of features misrepresent the information''} \cite{grinsztajn_et_al}.
\end{quote}

A second fundamental benefit of our approach is the method of learning the boolean function representation. Our proposed evolutionary scheme has the ability to bypass the local minima which may trap a traditional gradient-based tree boosting technique.

This paper proposes an alternative methodology to current ML and Deep Learning methods used in the prior art to make predictions from tabular data, and makes the following contributions:
\begin{enumerate}
    \item We establish a connection for the first time between circuit synthesis and a supervised ML problem via Graph-Based Genetic Programming, a form of evolutionary computing. No previous graph-based genetic programming research, to our knowledge, has considered a hardware circuit representation to be an ML predictor. 
    \item We propose a methodology called ``Auto Tiny Classifiers'' to automatically generate hardware circuits from tabular data for ML classification using graph-based genetic programming. Tiny Classifier circuits are composed of a very small number of logic gates (i.e., a few hundred) and are capable of matching the prediction accuracy of the state-of-the-art ML classifiers that are less efficient in area and power when implemented in hardware.
    \item We describe a toolflow that generates Tiny Classifiers as ASIC blocks. Then, we present the synthesis results of the Tiny Classifiers and ML baseline designs targeting the conventional Silicon technology. We also implement the Tiny Classifiers and ML baselines as FlexICs and fabricate them on flexible substrates (i.e., polyimide) using the flexible electronics fabrication technology. In addition, our toolflow generates Tiny Classifiers as Intellectual Property (IP) blocks so that they can be integrated as accelerators into a System-on-Chip (SoC). We demonstrate this in an Arm-based SoC with substantial FPGA resources in which the accelerators are synthesized.
\end{enumerate}

Tiny Classifiers can be used in many scenarios; e.g., triggering circuits within an SoC \cite{vocell}. A compelling scenario is to maintain an SoC in a low-power state while Tiny Classifiers are the always-on circuits. Once a situation of interest is uncovered by the classifier, then the rest (or subset) of the system would be awakened. Hardwired Tiny Classifiers can also be useful on their own for emerging Fast Moving Consumer Goods (FMCG) applications such as smart packaging enabled with flexible electronics (packages of dairy and meat products, labels of deodorant bottles, etc). Smart packages can be equipped with integrated circuits (ICs) fabricated on flexible substrates (e.g., plastic) using low-cost flexible electronics technology \cite{mubarik2020micro, weller2021date, biggs2021plasticarm}.

Flexible ICs are significantly less costly than Silicon-based ICs, paving the way to low-cost circuit customization \cite{bleier2022isca}. Tiny Classifiers can be implemented as flexible ICs closely coupled with low-cost printed sensors in a smart package and can make in-situ real-time predictions. There are recent examples of proposing and demonstrating ML models as flexible ICs to make in-situ classifications \cite{ozer2019fleps, ozer2020plasticML, armpit} in emerging FMCG products. These are typical near-sensor computing system \cite{zhou2020naturelectronics, iyerozer2016} examples where a compute block is closely coupled with a sensor, and the sensor data are turned into knowledge in the form of inference immediately at the source by low-cost and energy-efficient hardware. The programmability of classifier circuits is not a requirement for smart packages because of short FMCG product lifetimes (e.g., days or weeks) where products along with their packages will be disposed/recycled after use.

The remainder of the paper is organized as follows. Section \ref{sec:backgroundonEG} provides a brief introduction to graph-based genetic programming. 
Section \ref{sec:evolvingCC} discusses the adaptation of the evolutionary algorithm used in Auto Tiny Classifiers. Section \ref{sec:toolflow}  describes Auto Tiny Classifiers for generating Tiny Classifier circuits. Section \ref{sec:evaluation} describes the tabular datasets used in the experimental evaluation and shows the performance, hardware design and implementation results.
Finally, Section \ref{sec:related_work} presents the related work, and Section \ref{sec:conclusions} concludes the paper.

\section{Background}
\label{sec:backgroundonEG}

\subsection{Graph-based Genetic Programming}

\begin{figure}[h!]
    \centering
    \includegraphics[width=\linewidth]{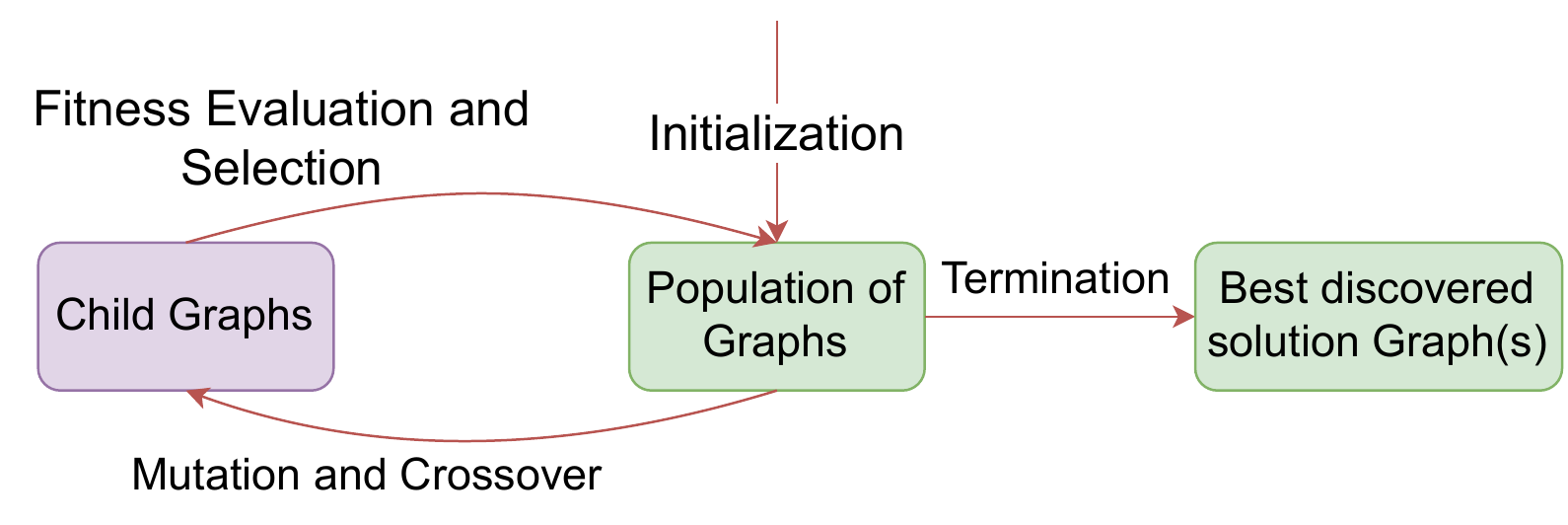}
    \caption{Overview of the Graph-based Genetic Programming methodology.}
    \label{fig:gbgp_overview}
\end{figure}

The general graph-based genetic programming approach \cite{miller2008cartesian, atkinson2018evolving, brameier2007linear, poli1997evolution}  follows a traditional evolutionary methodology (see Figure \ref{fig:gbgp_overview}). A set of possible solutions (the `\textit{population}') are recombined (`\textit{crossover}') and/or perturbed (`\textit{mutation}'). The new, candidate, solutions (the `\textit{children}') are then evaluated for their performance on the given task (giving a score, typically referred to as the `\textit{fitness}'). The best-performing children are selected to form the new population in the next iteration. Under the assumption that the problem has some sort of local continuity, such that children generated by performing crossover or mutation on high-quality solutions are more likely to be high-quality than randomly generated solutions, the algorithm tends towards higher-quality solutions over time. In doing so, it mimics natural Darwinian evolution, with the fitness acting as a selection pressure on the population, and mutation and crossover operators introducing variation.  

Graph-based genetic programming has been applied directly to both functional programs \cite{leitner2012mars,parziale2021cartesian} and stateful programs \cite{brameier2001evolving}. Graphs representing artificial neural networks have also been studied \cite{ahmad2012bio,turner2013cartesian}. The use of graph-based genetic programming for circuit synthesis has been considered in the literature \cite{miller1997designing,miller1999empirical, atkinson2018evolving,sotto2020study,franccoso2021graph}, where the most prominent technique, Cartesian Genetic Programming (CGP), rooted in circuit synthesis has remained a relevant benchmark task \cite{walker2008automatic,harding2011self,hodan2021semantically}. Such existing studies typically consider the task of synthesis against a completely known truth table, even when working with approximate circuit synthesis \cite{vasicek2014evolutionary,mrazek2017evoapprox8b}, where some error on that known truth table is acceptable in a tradeoff for greater efficiency. 


In contrast, only a fraction of the truth table is known in our ML setting (tabular data classification), and the population consists of circuits, represented as graphs, which are evaluated for their ability to correctly classify the training data. The final performance is measured with respect to the ability of the generated circuit to generalise as measured with the unseen test data.


\subsection{AutoML, NAS, NAIS versus Auto Tiny Classifiers}

\begin{figure}[!hbt]
    \centering
    \includegraphics[scale=0.51]{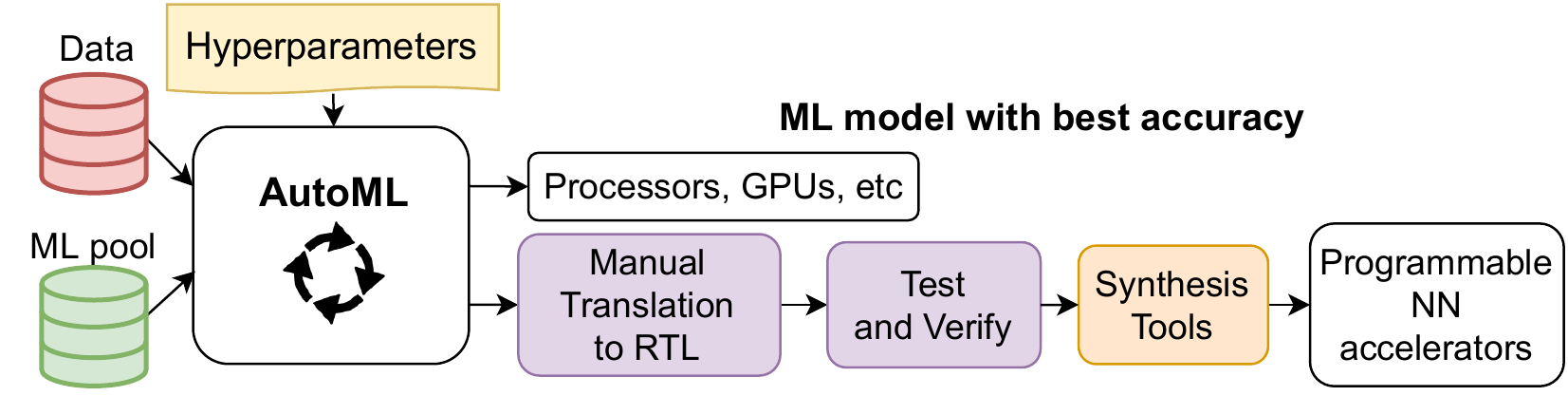}
    \caption{Automated Machine Learning (AutoML).}
    \label{fig:subfig_autoML}
\end{figure}

\begin{figure}[!hbth]
    \centering
    \includegraphics[scale=0.51]{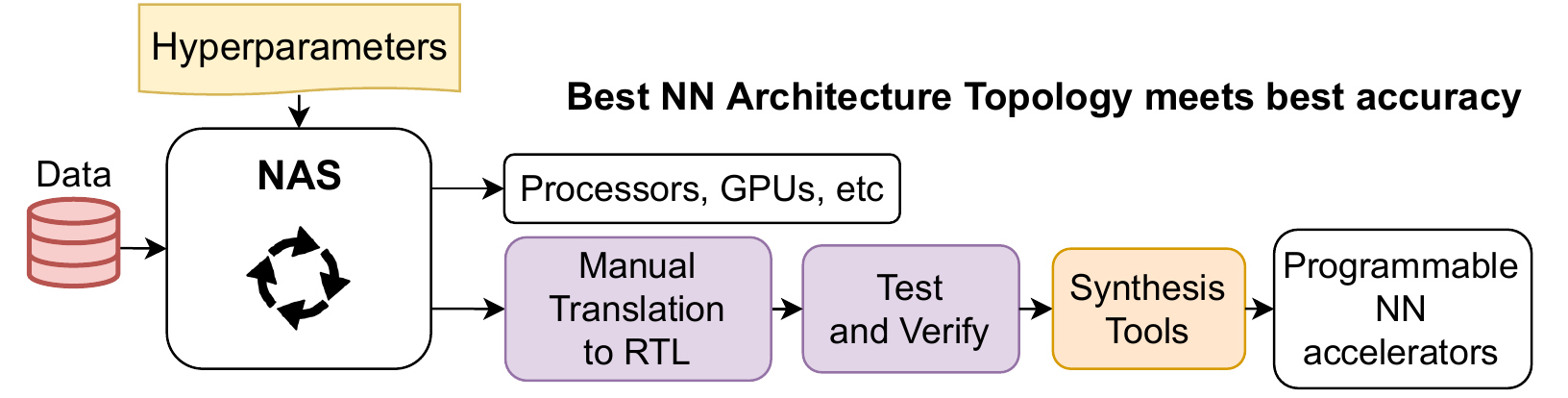}
    \caption{Neural Architecture Search (NAS).}
    \label{fig:subfig_nas}
\end{figure}

\begin{figure}[!hbt]
    \centering
    \includegraphics[scale=0.52]{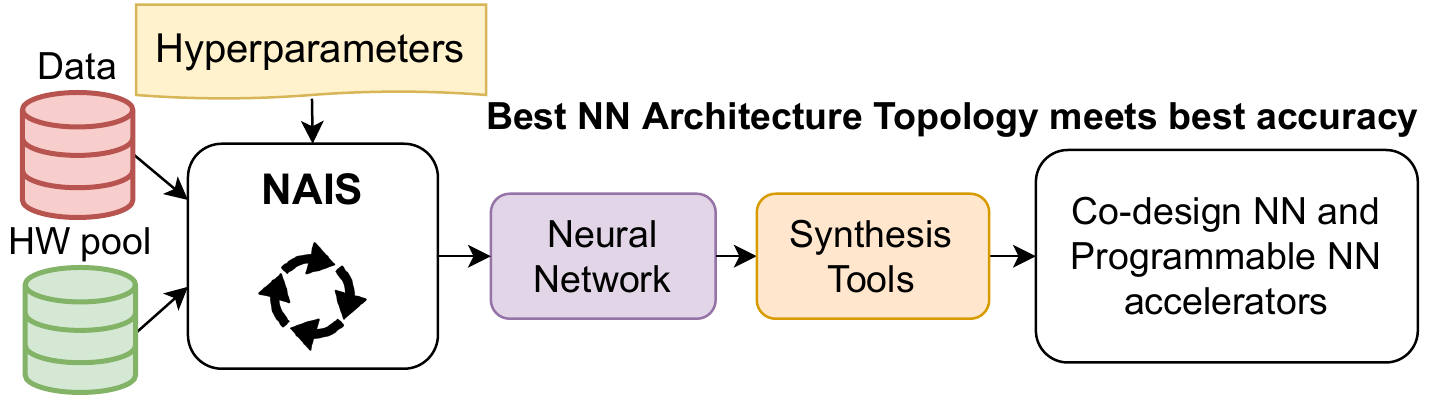}
    \caption{Neural Architecture and Implementation Search (NAIS).}
    \label{fig:subfig_nais}
\end{figure}

\begin{figure}[!hbt]
    \centering
    \includegraphics[scale=0.48]{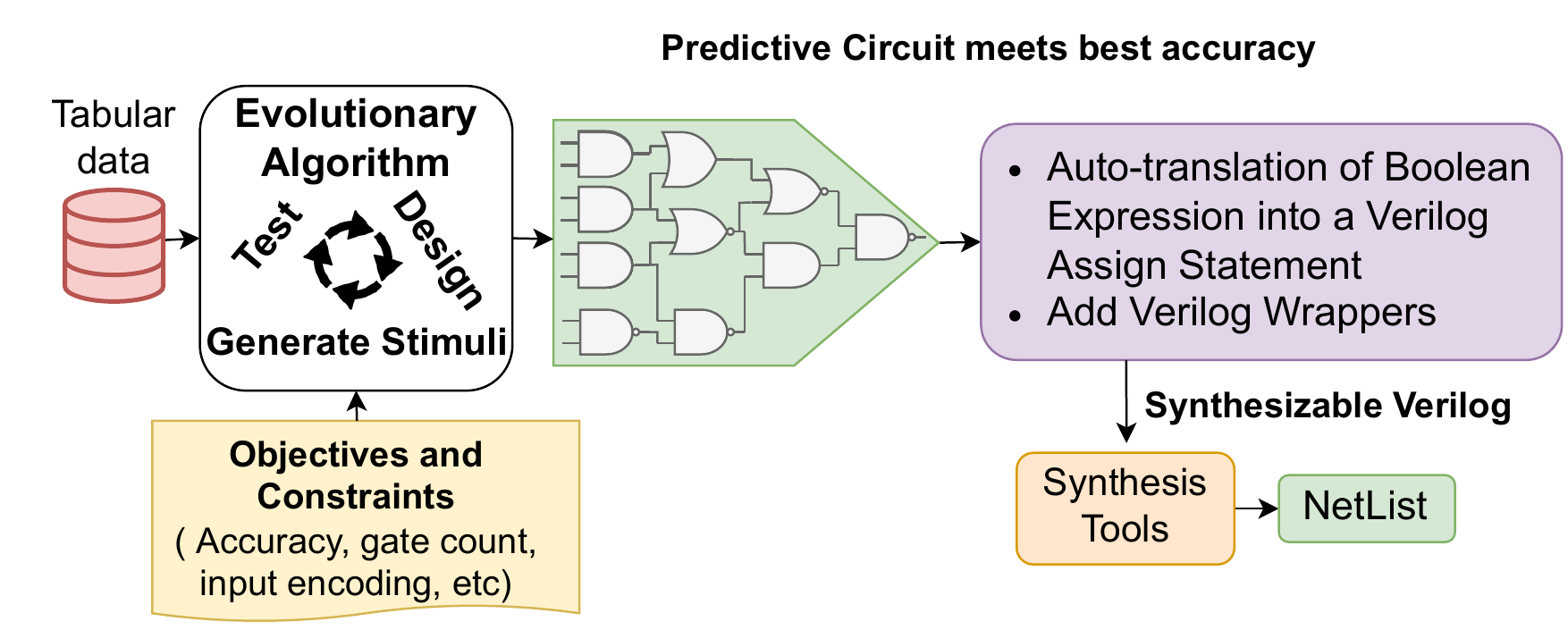}
    \caption{Auto Tiny Classifiers.}
    \label{fig:subfig_classifiers}
\end{figure}

Figures \ref{fig:subfig_autoML}, \ref{fig:subfig_nas}, \ref{fig:subfig_nais} and \ref{fig:subfig_classifiers} highlight the differences between current approaches of AutoML, Neural Architecture Search (NAS), Neural Architecture and Implementation Search (NAIS), and our Auto Tiny Classifier Circuits methodology for generating ML hardware as accelerators.

AutoML in Figure \ref{fig:subfig_autoML} and NAS in Figure \ref{fig:subfig_nas} generate an ML model and a Neural Architecture model, respectively, with maximised prediction performance. However, the ML model must be translated into RTL, which, in turn, still needs to be verified. NAIS, in Figure \ref{fig:subfig_nais}, selects a specific Neural Network and a known Neural Network accelerator to iterate over the space, identifying the best parameters from the hardware pool to maximise the prediction accuracy.

On the other hand, our proposed methodology searches the classifier circuit space automatically using an evolutionary algorithm, as shown in Figure \ref{fig:subfig_classifiers}. During circuit evolution, the generated circuit does not map to any predefined ML model or known hardware circuit. At the end of the search space cycle, the output is a sea of gates (a combinational circuit), which is auto-translated into RTL (i.e., typically as multiple Verilog assign statements for each classification output bit). These circuits are already verified during the fitness phase of the evolutionary algorithm. Our methodology is not a co-design approach, so there are no assumptions about any ML models or pre-determined hardware accelerator pools.

\section{Automatically Evolving Classifier Circuits}
\label{sec:evolvingCC}

Tabular data is addressed as a partial truth table of a circuit consisting of a sea of logic gates, where multiple heterogeneous features of the table are considered as the inputs to the circuit, whilst the classifier predictions are considered as its outputs. The fact that features in the tabular data are weakly correlated allows the conversion of the input-to-output prediction problem, into a simple representation of logic gates that can make predictions. 

We adapt the Evolving Graphs by Graph Programming (EGGP) algorithm \cite{atkinson2018evolving} as the evolutionary algorithm to generate the classification circuits. EGGP follows the consensus of using the simple $1+\lambda$ evolutionary technique \cite{miller2019cartesian}, and particularly for circuit synthesis  -- a practice reinforced by recent empirical experiments \cite{sotto2020study,franccoso2021graph}. The algorithm consists of the following steps:

\begin{enumerate}
    \item Generate a random initial parent solution $S$, and evaluate its fitness $f_S$.
    \item While not terminated do:
    \begin{enumerate}
        \item Generate $\lambda$ children $C_1 \dots C_\lambda$ by mutating $S$.
        \item Evaluate the children's fitness values $f_1 \dots f_\lambda$.
        \item If any child $C_i$ has $f_i \ge f_S$, then replace the parent $S = C_i$, $f_S = f_i$. Where multiple children satisfy this condition, the child with the highest fitness is chosen, tie-breaks are determined at random.
    \end{enumerate}
\end{enumerate}

The presence of the $\ge$ operator in the selection of a new parent, rather than just $>$, plays a pivotal role in the performance of the algorithm. In allowing the parent to be replaced by a child with equal fitness, the algorithm mimics the neutral drift of DNA as described in \cite{kimura1983neutral}. This allows the algorithm to undergo a `random walk' in the space of equivalent solutions, to the best solution so far, exposing the algorithm to new neighbourhoods of possible children and thereby allowing it to escape local optima. This simple modification yields significant performance gains in practice \cite{yu2001neutrality, yu2002finding,turner2015neutral} and may be augmented for further gains \cite{atkinson2019evolving,atkinson2020horizontal,downing2006neutrality,turner2015neutral}, although we do not use these extensions in this work. 

\subsection{Solution Representation}
\label{sec:sec:solution_representation}

In the algorithm, functional programs such as digital circuits are represented as graphs consisting of:

\begin{itemize}
    \item A set of input nodes $V_I$, each node of which uniquely represents an input to the program.
    \item A set of function nodes $V_F$, each node of which represents a specific function applied to its inputs.
    \item A set of output nodes $V_O$, each node of which uniquely represents an output of the program. 
    \item A set of edges $E$ connecting function nodes and output nodes to their respective inputs.
\end{itemize}

While in general the edges of each node are ordered so that they appropriately handle commutative functions \cite{atkinson2018evolving}, in this case, all considered functions are symmetric.

A crucial property of the EGGP representation is that function nodes need not be `active'. If there exists no path from a function node to an output node, then that node has no semantic meaning in the graph. This inactive material can be freely mutated to provide a direct mechanism for neutral drift.



\subsection{Genetic Operators}

When using the $1+\lambda$ evolutionary algorithm there are two main forms of genetic operator; initialisation and mutation.

\paragraph{Initialisation}

The initialisation is parameterized by the number of function nodes $n$, and the set of possible functions $F$. First, the $I$ input nodes $i_1 \dots i_I$ are created. Then for each $i \in 1 \dots n$, a function node $v_i$ is created and associated with a function chosen uniformly at random from $F$. $v_i$ is then connected uniformly at random to existing nodes $i_1 \dots i_I, v_1 \dots v_{i-1}$ until its degree matches the number of expected inputs to $f$. Finally, the $O$ output nodes $o_1 \dots o_O$ are created, and each is connected uniformly at random to a single node in $i_1 \dots i_I, v_1 \dots v_n$. The hyper-parameter $n$ determines the overall size of the graphs throughout the duration of the evolutionary run.

\paragraph{Mutation}

Mutation on solutions is performed via point mutations drawn from binomial distributions. The mutation rate $p$ parameterises the two binomial distributions $B(n, p)$ and $B(E, p)$ describing mutations of the functions nodes and edges, respectively. With $m_n \sim B(n, p)$ and $m_e \sim B(E, p)$ as the number of node and edge mutations to apply to the graph, the total $m_n + m_e$ mutations are applied in a randomly shuffled order, where;

\begin{itemize}
    \item For node mutations, a random function node $v \in V_f$ is chosen, and its associated function $f$ is replaced with $f' \in F, f' \neq f$ chosen uniformly at random. As the functions used here are symmetric and of the same arity, there is no need for input shuffling or connection modification procedures as described in \cite{atkinson2019evolving}.
    \item For edge mutations, a random edge $e \in E$ is chosen, where $s$ is the source of $e$ and $t$ is the target of $e$. The edge is redirected such that its new target $v \in V_I \bigcup V_F$ is chosen uniformly at random where the following conditions hold:
    \begin{itemize}
        \item There is no path $v \rightarrow s$ as this would introduce a cycle.
        \item $v \neq t$ as this would not introduce any perturbation of the solution. In the special (very rare) case that the number of inputs $I = 1$ and there is only a single node $t = i_1$ satisfying the first condition, the mutation is abandoned.
    \end{itemize}
\end{itemize}

\subsection{Fitness}
\label{sec:sec:fitness}

For all experiments performed here, the fitness of a circuit $C$ is its balanced accuracy. In general, other fitness functions could be supported, including additional objectives such as the number of gates or power consumption, which could be handled through the use of multi-objective graph-based genetic programming \cite{hilder2010use} to search for the Pareto-optimal front of solutions and characterize the trade-off between the objectives. In experiments performed here, the evolutionary algorithm simply attempts to maximize the accuracy for a given dataset and has no prior knowledge of what the eventual prediction accuracy of the classifier circuit should be. Our methodology offers the option to split the data into training and validation sets (with a 50-50\% split by default).
During evolution, the fitness of circuits is evaluated on both the training and validation set separately. The fitness of the training set determines the selection of children to replace the parent, whereas the fitness of the validation set ultimately determines the `best-discovered solution'. Effectively, we are maximising performance on the training set, while using the validation set to attempt to identify the best-generalised solution. The performance reported later in this paper is the performance on the reserved (unseen) testing set, as described in Section \ref{sec:evaluation}.



\subsection{Termination}
\label{sec:sec:termination}

In this setting, where the theoretical perfect accuracy of $100\%$ may never be achieved, we require a termination condition. We use a simple model, whereby if the validation fitness (computed on the 50\% validation set) has not improved by at least $\gamma$ within $\kappa$ generations, the algorithm terminates and returns the best-discovered solution with respect to the validation data. Additionally, the algorithm will automatically terminate if the number of generations exceeds the threshold $G$.

\subsection{Hyperparameters}
\label{sec:sec:hyperparameters}

The hyperparameters of the algorithm are as follows:

\begin{itemize}
    \item The number of children per generation, $\lambda$.
    \item The mutation rate, $p$.
    \item The function set from which solutions may be constructed, $F$.
    \item The termination threshold $\gamma$.
    \item The corresponding window of generations to achieve that threshold and terminate, $\kappa$.
    \item The maximum number of generations $G$.
\end{itemize}

In Section \ref{sec:sec:design_space_exploration} we vary the function set $F$, number of function nodes $n$, termination generations $\kappa$ and maximum number of generations $G$ to choose hyper-parameters for evaluation in Section \ref{sec:accuracy_comparison}. The other hyper-parameters use the fixed values: $\lambda=4$, $p=\frac{1}{n}$, $\gamma = 0.01$.

%
%

\subsection{Classifier Circuits as Accelerators}
\label{sec:sec:classifiers_accelerators}

The system can be thought as a set of classification circuit block(s) or a single classification circuit unit which lead to classification ``guesses''. The prediction could be a single bit (binary classification) or a set of bits in the case of multiclass classification problems which represent the encoding of the target class. Except for the actual classification circuit, the design uses buffers to hold the input and output data. The use of local buffers eliminates the data transfers within the system, keeping the required data close to the computation block(s).

\begin{figure}[!tbh]
     \centering
     \includegraphics[width=0.48\textwidth]{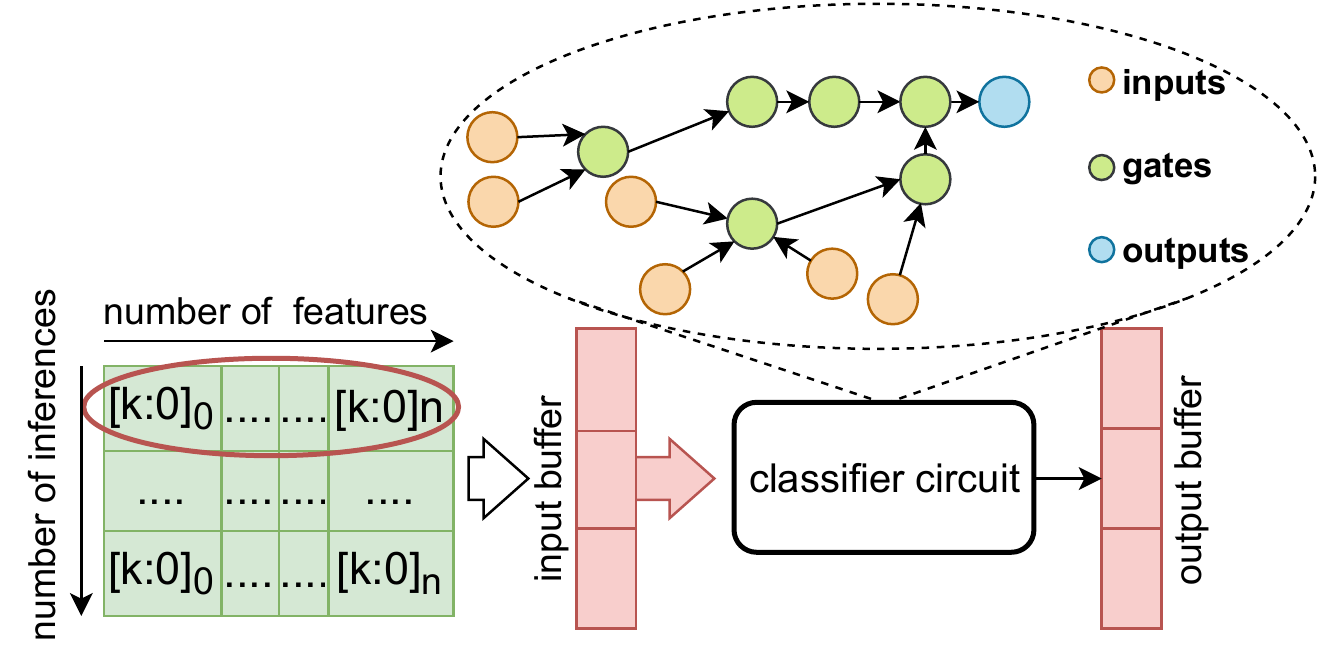}
     \caption{One instance of a classifier circuit.}
     \label{fig:one_instance}
\end{figure}
 


Figure \ref{fig:one_instance} presents classifier circuits as accelerators within a system. The inputs of the classifier circuit are single bits. The number of inputs for one classification circuit can be defined as the $ number\_of\_features\_in\_one\_inference \times encoding\_bits\_per\_input$. Each feature of the inference is transformed into a group of bits based on the input encoding and the preferred number of bits per input. These parameters are user-defined. Most of the classification circuits use only a subset of input bits to perform a prediction. As a result, the above expression is the upper bound for an input size buffer. The actual size of the local buffer is determined after the generation of the classification circuit and it holds only the necessary bits which will be consumed by the classification circuit for the prediction.

In the case of binary classification where the prediction is `0' or `1' (`yes' or `no'), the output of the classifier is one bit. Basically, for each inference, we produce one classification and the result (single bit) is placed in the output buffer. However, for multiclass classification problems, the classification circuits have more than one output, which indicate the encoded predicted class. As a result, we instantiate $bits\_per\_output$ (user-defined parameter) local output buffers, which hold the encoded prediction for every inference. Of course, the size and the number of local input/output buffers increase the ``cost'' of the accelerator and this tradeoff should be explored based on the hardware specifications of the target embedded system.

Figure \ref{fig:one_instance} presents the simplest accelerator which evolves classification circuits. It is the smallest possible accelerator design which includes a classification circuit. Identical classification circuits can be combined and process multiple inferences in parallel.  
In that case, the number of input local buffers is the number of parallel classification circuits within the accelerator. The processing of multiple inferences can be done in parallel, as long as there are available resources.

\section{Auto Tiny Classifiers}
\label{sec:toolflow}
\begin{figure*}[thb!]
\centering
    \includegraphics[width=\textwidth]{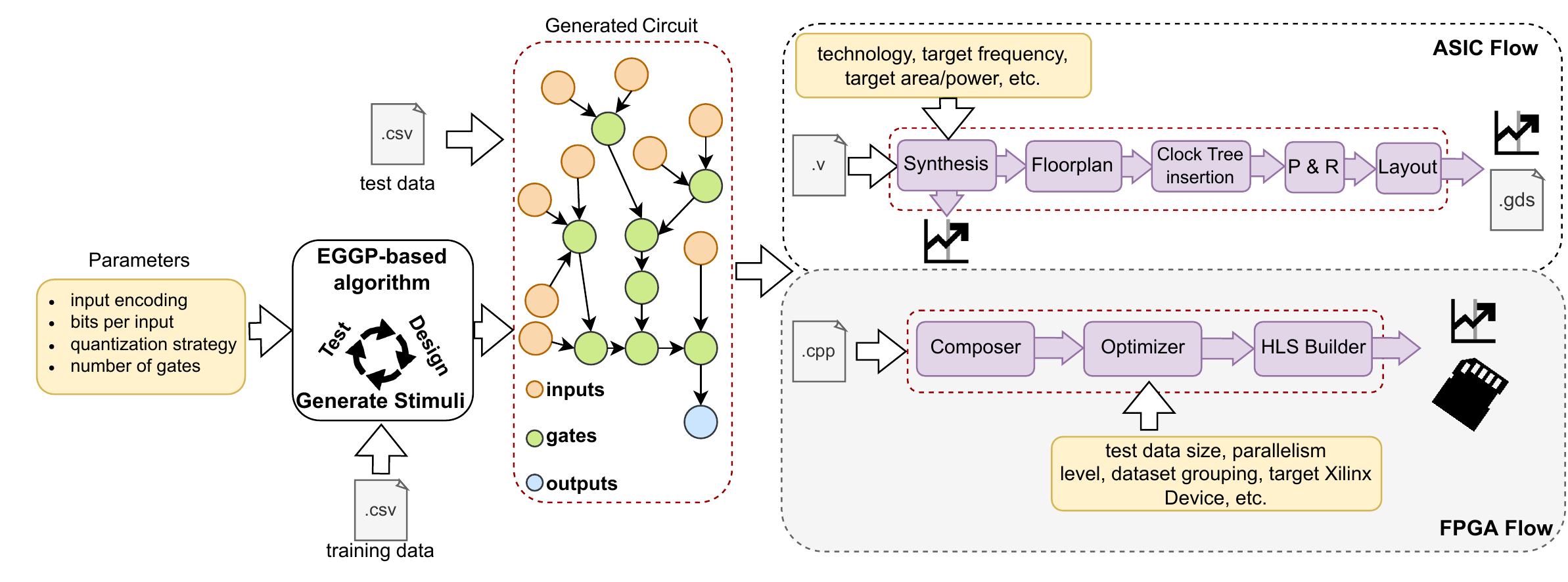}
    \caption{Methodology for generating Tiny Classifier circuits.}
    \label{fig:toolflow}
\end{figure*}



Figure \ref{fig:toolflow} shows the methodology of automatically generating Tiny Classifier circuits as hardware accelerators targeting ASIC and FPGA platforms.

\subsection{From Tabular Data to Circuit Representations}

The proposed methodology generates a visual representation of the classifier circuit directly from the training data and user-defined input parameters, as shown in Figure \ref{fig:toolflow}. The input parameters can be a subset or full set of the following; the total gate count of the classifier circuits, the type of the input encoding (binary, one-hot, gray), the number of required bits per input for the encoding and the quantization strategy (quantization/quantiles). The EGGP-based evolutionary algorithm crawls on the design space using the training data and converges on a simple graph of a sea of logic gates as the output circuit representation on which the test data is used to measure the final prediction accuracy. The sea of logic gates is automatically translated into RTL (e.g., Verilog).

\subsection{ASIC and FPGA Target}
\label{sec:sec:fpga-asic-target}

Auto Tiny Classifiers generate Tiny Classifier circuits that can be implemented in ASIC, as shown in the dotted box inside the ``ASIC flow'' of Figure \ref{fig:toolflow}. The auto-generated Verilog representation of a Tiny Classifier is read by the synthesis tool that generates the netlist for a given technology standard cell library and constraints, and then produces the synthesized area, power and timing reports. The full chip implementation requires steps beyond synthesis such as floorplanning, clock tree insertion, place \& route, and layout rules checking and generation. The output of the flow is the generation of the chip layouts in GDS format to complete the tape-out as well as the area, power and timing reports of the full implementation.

To target FPGAs, we use Xilinx SDSoC which is a software/hardware development environment from Xilinx for Zynq platforms \cite{xilinx_sdsoc_env_user} \cite{xilinx_sdsoc_prog_guide}. Xilinx SDSoC offers a complete software/hardware co-design environment where the designers develop the system software part along with a C/C++ implementation of the accelerated function, which will then be compiled by High-Level Synthesis (HLS). All layers between the host application and the hardware RTL (drivers, Operating Systems, etc.) are provided through an automated process. For our experiments, we use Xilinx Zynq Ultrascale+ MPSoC as a target platform which consists of a low-power processing system (PS) with a quad-core Arm Cortex-A53 coupled with a user-programmable logic part (PL). The generated Tiny Classifier circuit is the accelerator IP, which is then translated into C/C++ code for High-Level Synthesis to target an SoC platform partitioning the code between an FPGA target and a CPU \footnote{It is also possible to use the generated Verilog module from our tool-flow to create a software/hardware co-design solution. However, all the layers (drivers, OS, etc.) must be developed manually without the automated process from Xilinx tools.}. After the transformation of a classifier circuit to a C/C++ function, the generated circuit is ready to be synthesized. This process is separated into three main phases: (a) Composer - gathers information about the generated classifier circuit and creates the necessary project files for the Xilinx SDSoC tools, (b) Optimizer - optimizes the Xilinx SDSoC generated project (\texttt{\#pragma} directives for the HLS compiler, data transfer configuration between the PS and PL part of the target platform, etc) and (c) HLS Builder - produces a ready plug-and-play image with an integrated lightweight OS for the target platform including all the necessary libraries in a software/hardware co-design environment.

\section {Evaluation}
\label{sec:evaluation}


The experiments use a comprehensive collection of 33 tabular datasets, mainly from OpenML \cite{OpenML2013, OpenML2017, OpenML2020}, UCI \cite{uci_repository} and Kaggle \cite{kaggle}. For the datasets, we select the 17 used by Kadra \textit{et al} which serve as a representative benchmark collection for tabular data \cite{regularization_is_all_you_need}, and in addition, focus on 16 mainly multiclass classification datasets from UCI, Kaggle and OpenML. For example, the dataset \emph{higgs} contains sensor data from high-energy physics \cite{higgs}. The dataset \emph{clickpred} contains advertisements in a search engine, and whether or not they were clicked. From the selected collection, 14 tabular datasets were used by AutoGluon Tabular \cite{agtabular}, the state-of-the-art AutoML tool for tabular data.

\begin{table}[h!]
\centering
\begin{threeparttable}
\begin{tabular}{llll}
\hline
\multicolumn{1}{|l|}{Dataset (Source)}  & \multicolumn{1}{l|}{Classes} & \multicolumn{1}{l|}{Rows} & \multicolumn{1}{l|}{Features} \\
\hline
\hline
\multicolumn{1}{|l|}{\dag vehicle (OpenML)}          & \multicolumn{1}{l|}{2}    & \multicolumn{1}{l|}{846}      & \multicolumn{1}{l|}{22}   \\ \hline
\multicolumn{1}{|l|}{\dag cars (OpenML)}             & \multicolumn{1}{l|}{3}    & \multicolumn{1}{l|}{406}      & \multicolumn{1}{l|}{8}    \\ \hline
\multicolumn{1}{|l|}{user model data (UCI)}      & \multicolumn{1}{l|}{4}    & \multicolumn{1}{l|}{403}      & \multicolumn{1}{l|}{5}    \\ \hline
\multicolumn{1}{|l|}{\dag kc1 (OpenML)}              & \multicolumn{1}{l|}{2}    & \multicolumn{1}{l|}{145}      & \multicolumn{1}{l|}{95}   \\ \hline
\multicolumn{1}{|l|}{\dag phoneme (OpenML)}          & \multicolumn{1}{l|}{2}    & \multicolumn{1}{l|}{5404}     & \multicolumn{1}{l|}{6}    \\ \hline
\multicolumn{1}{|l|}{skin-seg (OpenML)}         & \multicolumn{1}{l|}{2}    & \multicolumn{1}{l|}{245057}   & \multicolumn{1}{l|}{4}    \\ \hline
\multicolumn{1}{|l|}{ecoli-data (UCI)}          & \multicolumn{1}{l|}{4}    & \multicolumn{1}{l|}{336}      & \multicolumn{1}{l|}{8}    \\ \hline
\multicolumn{1}{|l|}{iris (UCI)}                & \multicolumn{1}{l|}{3}    & \multicolumn{1}{l|}{150}      & \multicolumn{1}{l|}{7}    \\ \hline
\multicolumn{1}{|l|}{\dag blood (OpenML)}            & \multicolumn{1}{l|}{2}    & \multicolumn{1}{l|}{748}      & \multicolumn{1}{l|}{4}    \\ \hline
\multicolumn{1}{|l|}{\dag higgs (OpenML)}            & \multicolumn{1}{l|}{2}    & \multicolumn{1}{l|}{98050}    & \multicolumn{1}{l|}{29}   \\ \hline
\multicolumn{1}{|l|}{wifi-localization (UCI)}   & \multicolumn{1}{l|}{4}    & \multicolumn{1}{l|}{2000}     & \multicolumn{1}{l|}{7}    \\ \hline
\multicolumn{1}{|l|}{\dag nomao (OpenML)}            & \multicolumn{1}{l|}{2}    & \multicolumn{1}{l|}{34465}    & \multicolumn{1}{l|}{119}  \\ \hline
\multicolumn{1}{|l|}{olinda-outlier (OpenML)} & \multicolumn{1}{l|}{4}    & \multicolumn{1}{l|}{75}       & \multicolumn{1}{l|}{3}    \\ \hline
\multicolumn{1}{|l|}{\dag australian (OpenML)}       & \multicolumn{1}{l|}{2}    & \multicolumn{1}{l|}{690}      & \multicolumn{1}{l|}{15}   \\ \hline
\multicolumn{1}{|l|}{\dag segment (OpenML)}          & \multicolumn{1}{l|}{2}    & \multicolumn{1}{l|}{2310}     & \multicolumn{1}{l|}{20}   \\ \hline
\multicolumn{1}{|l|}{led (UCI)}                 & \multicolumn{1}{l|}{10}   & \multicolumn{1}{l|}{500}      & \multicolumn{1}{l|}{7}    \\ \hline
\multicolumn{1}{|l|}{\dag numerai (OpenML)}          & \multicolumn{1}{l|}{2}    & \multicolumn{1}{l|}{96320}    & \multicolumn{1}{l|}{22}   \\ \hline
\multicolumn{1}{|l|}{\dag miniboone (OpenML)}        & \multicolumn{1}{l|}{2}    & \multicolumn{1}{l|}{130064}   & \multicolumn{1}{l|}{51}   \\ \hline
\multicolumn{1}{|l|}{wall-robot (Kaggle)}       & \multicolumn{1}{l|}{4}    & \multicolumn{1}{l|}{5456}     & \multicolumn{1}{l|}{3}    \\ \hline
\multicolumn{1}{|l|}{\dag jasmine (OpenML)}          & \multicolumn{1}{l|}{2}    & \multicolumn{1}{l|}{2984}     & \multicolumn{1}{l|}{145}  \\ \hline
\multicolumn{1}{|l|}{yeast (UCI)}               & \multicolumn{1}{l|}{10}   & \multicolumn{1}{l|}{1484}     & \multicolumn{1}{l|}{8}    \\ \hline
\multicolumn{1}{|l|}{\dag christine (OpenML)}        & \multicolumn{1}{l|}{2}    & \multicolumn{1}{l|}{5418}     & \multicolumn{1}{l|}{1637} \\ \hline
\multicolumn{1}{|l|}{\dag sylvine (OpenML)}           & \multicolumn{1}{l|}{2}    & \multicolumn{1}{l|}{5124}     & \multicolumn{1}{l|}{21}   \\ \hline
\multicolumn{1}{|l|}{seismic-bumps (UCI)}       & \multicolumn{1}{l|}{3}    & \multicolumn{1}{l|}{210}      & \multicolumn{1}{l|}{8}    \\ \hline
\multicolumn{1}{|l|}{ccfraud (OpenML)}          & \multicolumn{1}{l|}{2}    & \multicolumn{1}{l|}{284807}   & \multicolumn{1}{l|}{31}   \\ \hline
\multicolumn{1}{|l|}{clickpred (OpenML)}        & \multicolumn{1}{l|}{2}    & \multicolumn{1}{l|}{1496391}  & \multicolumn{1}{l|}{10}   \\ \hline
\multicolumn{1}{|l|}{vowel (UCI)}               & \multicolumn{1}{l|}{2}    & \multicolumn{1}{l|}{528}      & \multicolumn{1}{l|}{21}   \\ \hline
\multicolumn{1}{|l|}{nursery (UCI)}             & \multicolumn{1}{l|}{5}    & \multicolumn{1}{l|}{12958}    & \multicolumn{1}{l|}{9}    \\ \hline
\multicolumn{1}{|l|}{spectf-data (Kaggle)}      & \multicolumn{1}{l|}{2}    & \multicolumn{1}{l|}{267}      & \multicolumn{1}{l|}{45}   \\ \hline
\multicolumn{1}{|l|}{teaching assist (UCI)}  & \multicolumn{1}{l|}{3}    & \multicolumn{1}{l|}{151}      & \multicolumn{1}{l|}{7}    \\ \hline
\multicolumn{1}{|l|}{wisconsin (UCI)}           & \multicolumn{1}{l|}{2}    & \multicolumn{1}{l|}{194}      & \multicolumn{1}{l|}{33}   \\ \hline
\multicolumn{1}{|l|}{sonar (Kaggle)}            & \multicolumn{1}{l|}{2}    & \multicolumn{1}{l|}{208}      & \multicolumn{1}{l|}{61}   \\ \hline
\multicolumn{1}{|l|}{ionosphere (UCI)}          & \multicolumn{1}{l|}{2}    & \multicolumn{1}{l|}{351}      & \multicolumn{1}{l|}{35}   \\ \hline
\end{tabular}

\begin{tablenotes}
   \item Note: \dag indicates that the dataset was appeared in the AutoGluon Tabular paper \cite{agtabular}.
\end{tablenotes}
\end{threeparttable}
\caption{The collection of the datasets.} 
\label{tab:datasets}
\end{table}

Table \ref{tab:datasets} provides the full list of datasets and their main characteristics. Each dataset is split into 80\% training and 20\% testing sets. The prediction accuracy results for both Tiny Classifier circuits and ML baseline models in the following subsections are based on test datasets.

\subsection{Selection of Baseline ML Models}
\label{sec:sec:baselines}

We use Google's TabNet DNN \cite{tabnet} with the recommended hyperparameters configuration, and AutoGluon (An AutoML system developed by Amazon) \cite{agtabular, agtabulardistill} with explicit support for tabular data (Tabular Predictor) as well as other baseline ML models. Google's TabNet is one of the first successful DNNs addressing tabular data, using sequential attention to select features for decision-making layers. AutoGluon searches the design space over three state-of-the-art models (i.e, XGBoost, TabNeuralNet and NNFastAITab) for Tabular Data among others. AutoGluon XGBoost is based on Gradient Boosting, whereas the other two models are based on DNNs. In our experiments, AutoGluon Tabular Predictor is configured with the above three models. \textit{Kadra et al}.\ \cite{regularization_is_all_you_need} observe that a Neural Architecture Search (NAS) over Multi-layer Perceptrons (MLPs) delivers state-of-the-art NN models for tabular data. Hence, we also use the NAS-based protocol described by \textit{Kadra et al}.\ \cite{regularization_is_all_you_need} to generate baseline MLP models.

\begin{figure*}[h!]
    \centering
    \includegraphics[width=0.9\textwidth]{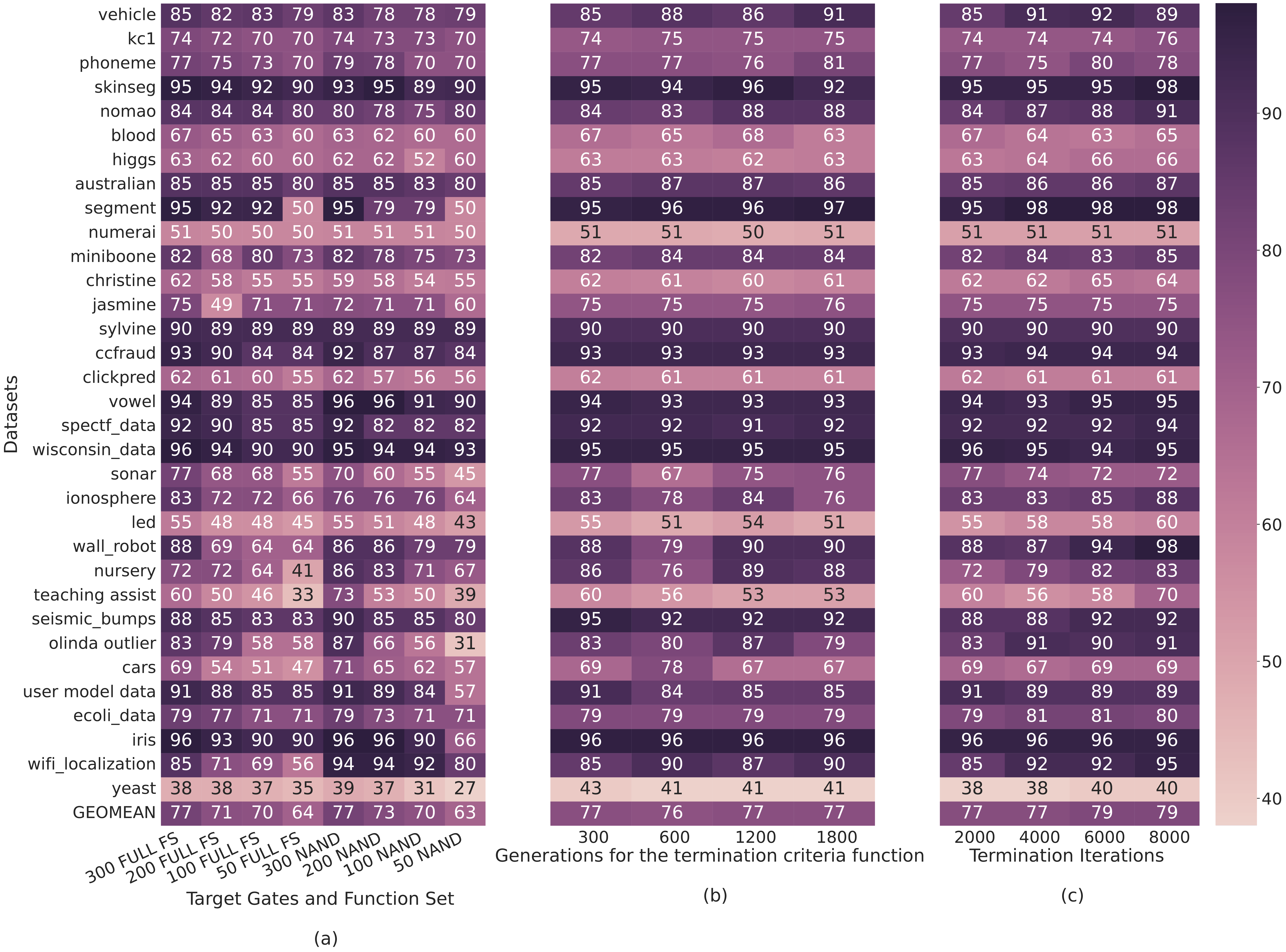}
    \caption{(a) Accuracy vs.\ number of gates. Generations for the termination function is 300 and termination iterations is 2000. \textit{Full FS} indicates that the generated circuit will be constructed with logical gates within the function set $F = \{\texttt{and}, \texttt{or}, \texttt{nand}, \texttt{nor}\}$. For NAND function set the generated circuits constructed only with NAND gates. (b) Accuracy vs.\ generations for the termination function. The number of gates is 300 and the number of termination iterations is 2000. (c) Accuracy vs.\ the number of termination iterations. The number of gates and the generations for the termination function are both set to 300.}
    \label{plot:exploration_heatmap}
\end{figure*}

\subsection{Data Encoding and Quantization Strategy}
\label{sec:sec:input-encoding-quantization}

 Numerical inputs are automatically handled to encode the features of a dataset based on user preferences. The encoding consists of the encoding strategy and the number of bits per input. The encoding strategy determines the way that numerical features get translated into binary. Currently, four main encoding strategies are supported: (a) \textit{quantization}, where each feature is divided into buckets of equal width, (b) \textit{quantiles}, where each feature is divided into buckets of width roughly equal numbers to the number of samples, (c) one-hot and (d) gray. Additionally, the users can manually tune the \textit{number of bits per input} to decide the granularity of the input encoding. From now onwards, experiments report only the best-achieved accuracy across the available encoding strategies with two and four bits per input.

In the comparative analysis with Tiny Classifiers, MLP models are transformed into 2-bit quantized versions. Since the hardware requirements of Tiny Classifiers are minimal, a comparison against the non-quantized MLPs does not provide a fair baseline when considering latency, area and power. Thus, we use a 2-bit quantized MLP as the resource-optimized high-performing baseline.

\subsection{Tiny Classifier Design Space}
\label{sec:sec:design_space_exploration}

A primary goal is to check whether we can generate accurate combinational logic for an ML classification problem. We explore different combinations of the hyperparameters (see Section \ref{sec:sec:hyperparameters}) of the evolutionary algorithm to improve the accuracy of the generated circuits for all the datasets.  Next, we explore the design space in four main directions: (a) the size of the generated circuit $n$ (number of gates), (b) the function set $F$ from which solutions (circuits) may be constructed, (c) the number of generations for the termination criterion function $\kappa$, and (d) the number of iterations to achieve a performance threshold and terminate $G$. 

The heatmap of Figure \ref{plot:exploration_heatmap}a presents the achieved accuracy of the generated Tiny Classifier circuits as we progressively decrease the target NAND gate count from 300 to 50. At the same time, we explore the accuracy of the circuits with two different function sets. Overall, we observe a 14 percentage points reduction in GEOMEAN across all datasets from 300 gates to 50 gates.

The next step is to study how the number of generations for the termination criterion function impacts the accuracy of Tiny Classifiers when we limit the circuit size to a maximum of 300 gates. Figure \ref{plot:exploration_heatmap}b shows the achieved accuracy for various generation values of the termination criterion function. No significant change in prediction performance is observed.

Figure \ref{plot:exploration_heatmap}c presents the number of termination iterations versus achieved accuracy. We progressively increase the number of termination iterations as we set the target gate count and the number of generations for the termination function to 300. We observe a 2 percentage points improvement in GEOMEAN accuracy across all datasets when increasing the number of iterations.

\begin{figure*}[h!]
    \centering
    \includegraphics[width=\textwidth]{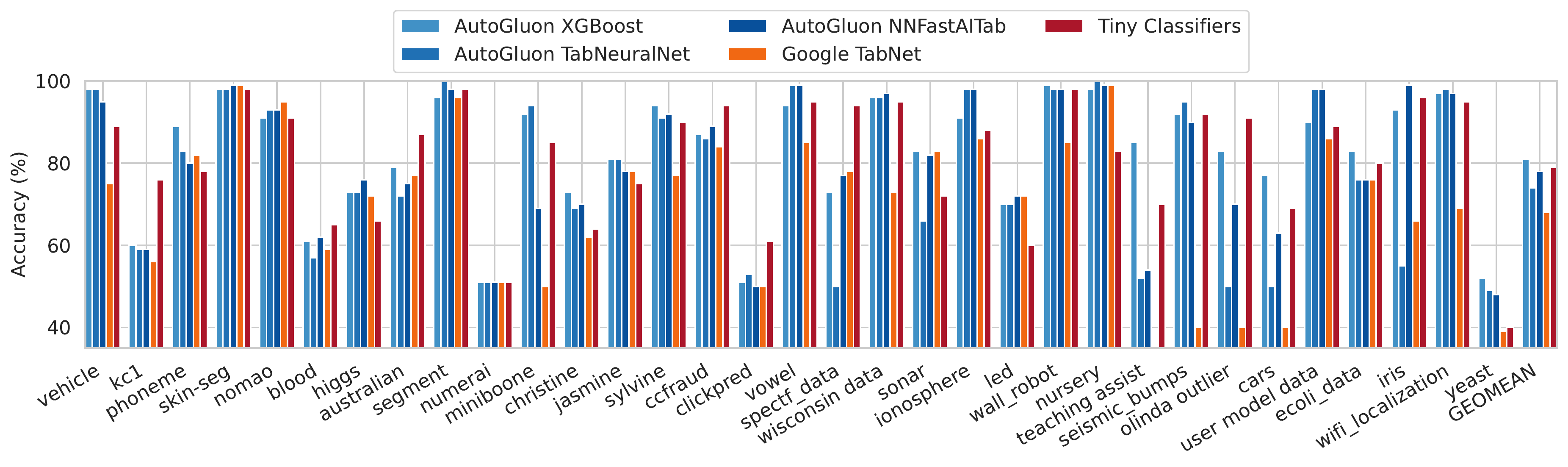}
    \caption{Prediction Accuracy of Tiny Classifiers, AutoGluon XGBoost, AutoGluon TabularNeuralNet, AutoGluon NNFastAITabular and Google TabNet. Note the datasets \emph{vehicle} to \emph{ionosphere} (left to right) are binary, and the remainder are multiclass classifications.}
    \label{plot:accuracy_pred_circuits_vs_autogluon}
\end{figure*}

\subsection{Accuracy Comparison}
\label{sec:accuracy_comparison}

\begin{figure*}[h!] 
    \centering
    \includegraphics[width=\textwidth]{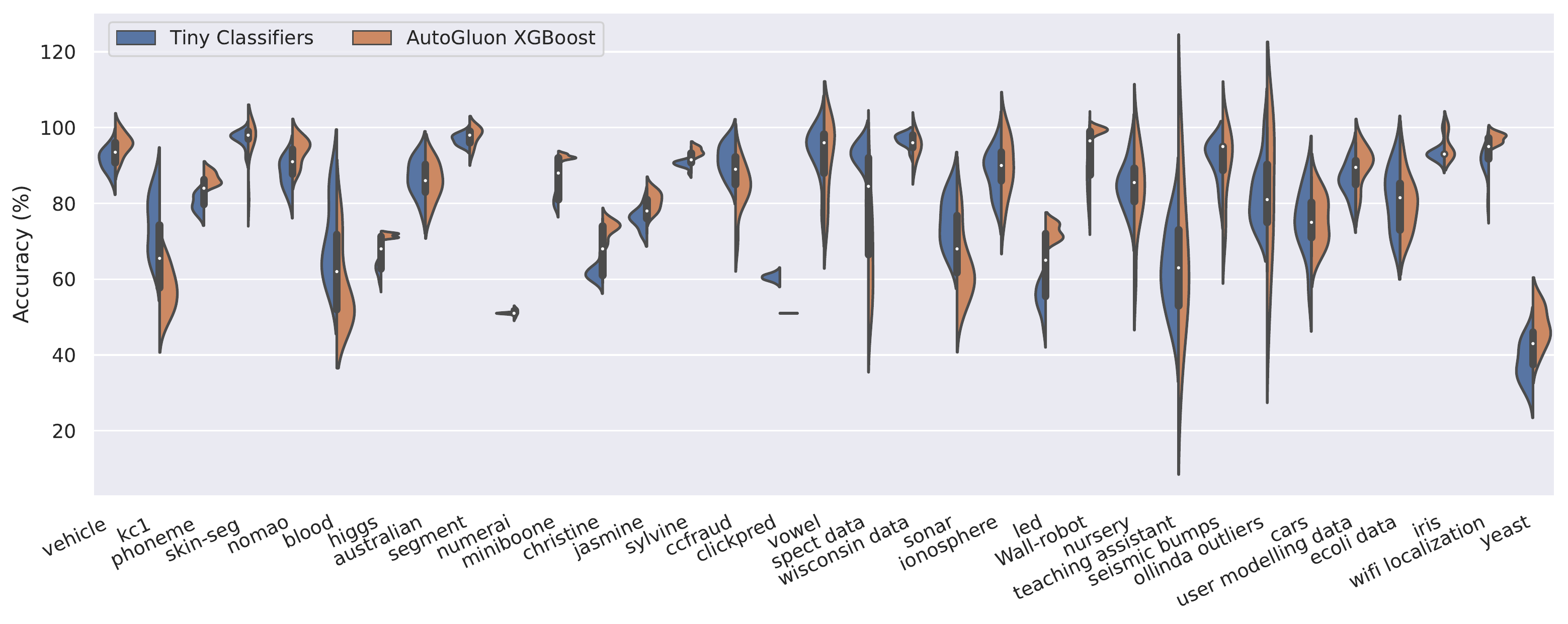}
    \caption{Violin plots showing the accuracy distributions of Tiny Classifiers and AutoGluon XGBoost.}
    \label{plot:robustness}
\end{figure*}

\begin{figure*}[h!] 
    \centering
    \includegraphics[width=\textwidth]{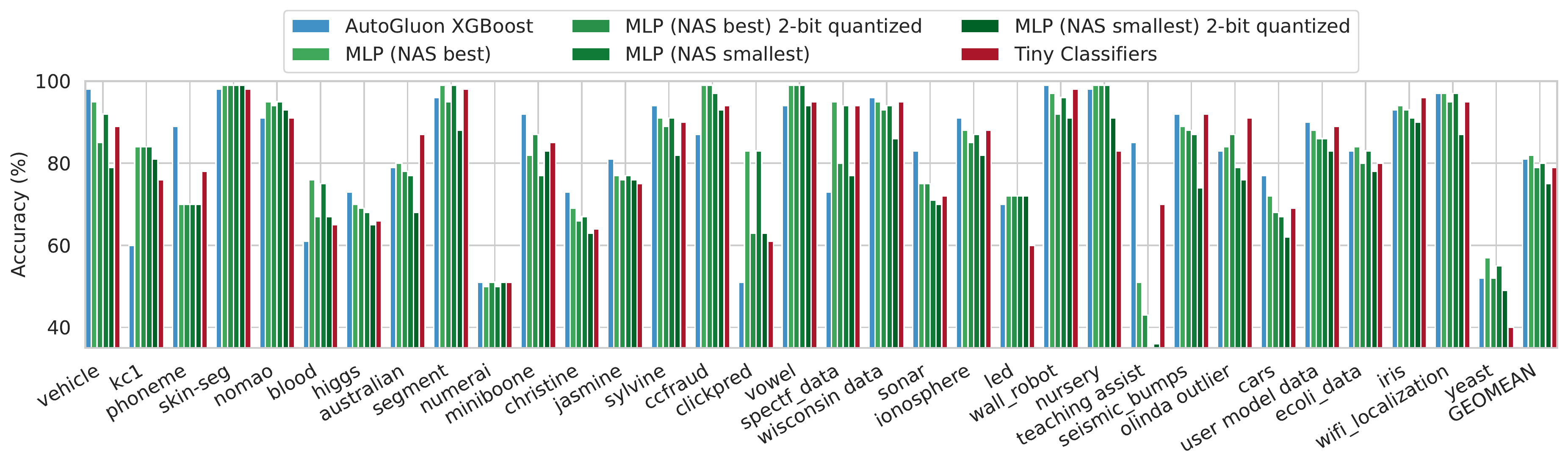}
    \caption{Prediction Accuracy of Tiny Classifiers, smallest MLP (non-quantized and 2-bit quantized versions), best MLP (non-quantized and 2-bit quantized versions) and AutoGluon XGBoost.}
    \label{plot:accuracy_pred_circuits_vs_mlp}
\end{figure*}

Figure \ref{plot:accuracy_pred_circuits_vs_autogluon} compares the prediction accuracy of Google TabNet, AutoGluon and Tiny Classifiers. Based on the analysis in Section \ref{sec:sec:design_space_exploration}, the hyperparameters of Tiny Classifiers are set to 300 for both the number of gates and the termination function. In addition, the maximum number of iterations is set to 8000. Across all the datasets, the average prediction accuracy of AutoGluon XGBoost is 81\%, which is the overall highest. The mean accuracy of Tiny Classifiers across all the datasets is 78\%, which is the second highest.

We compare the prediction accuracy distribution of Tiny Classifiers against AutoGluon XGBoost to understand how robust Tiny Classifiers are with respect to XGBoost. To this end, we perform a 10-fold cross-validation study and show the accuracy distributions of Tiny Classifiers and XGBoost in Figure \ref{plot:robustness} using a violin plot. 

The interquartile range of Tiny Classifiers is comparable to the interquartile ranges of ML baselines and in some cases, even slightly shorter. The shape of the distribution in Tiny Classifiers indicates that the accuracy data are highly concentrated around the median. This implies a low variance of the accuracy distribution and therefore makes Tiny Classifiers robust to variation.

\begin{figure*}[h!] 
    \centering
    \includegraphics[width=0.95\textwidth]{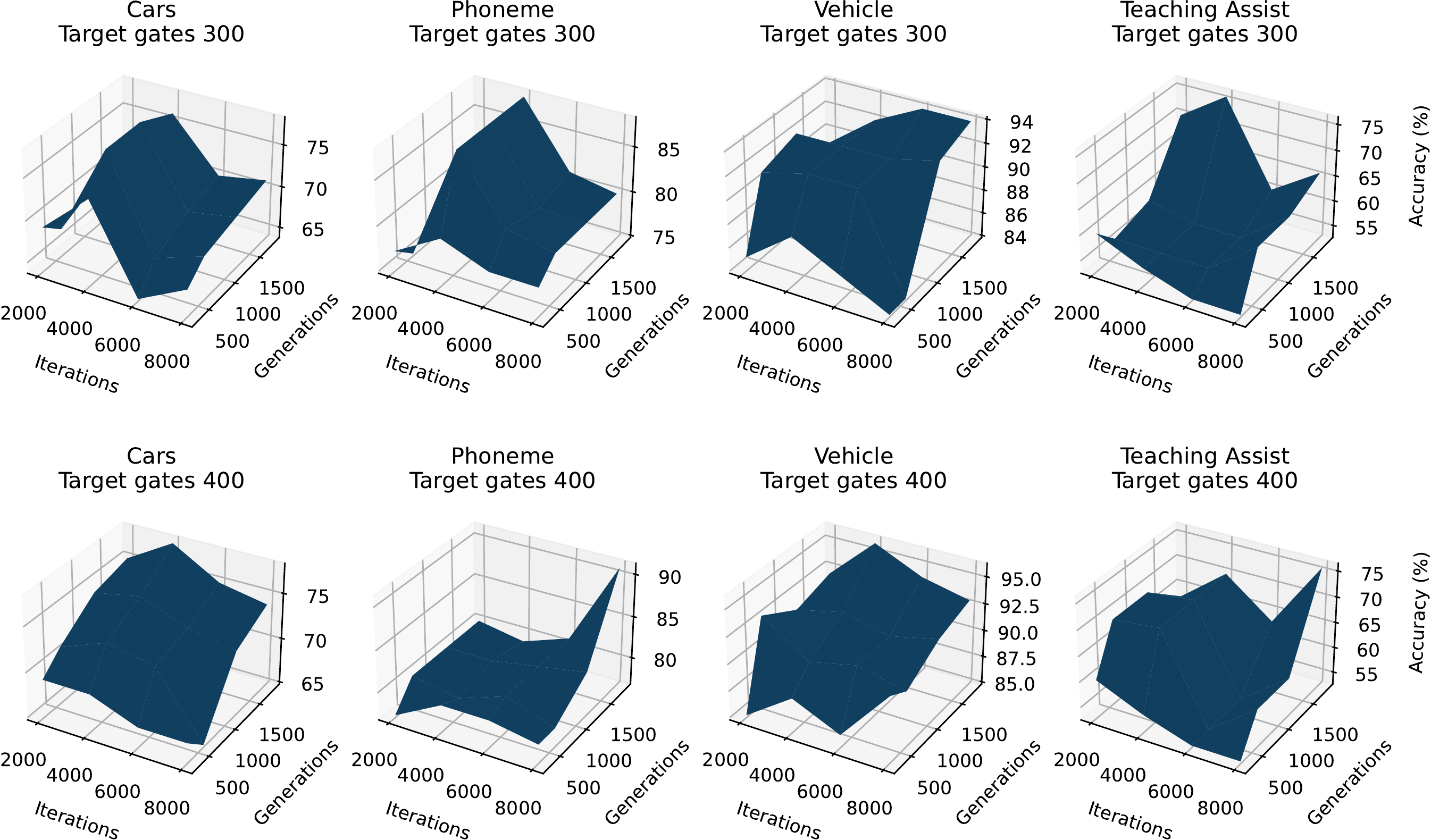}
    \caption{Space Exploration for four datasets (\textit{vehicle, phoneme, Teaching Assist and Cars}) beyond the limit of 300 logical gates. The best achieved prediction accuracy of Tiny Classifier Circuits with 400 gates ranges from 76\% - 96\%.}
    \label{plot:further_exploration}
\end{figure*}

The best-performing ML model, XGBoost, and Tiny Classifiers from Figure \ref{plot:accuracy_pred_circuits_vs_autogluon} are also compared to the best and smallest MLP configurations. We first explore the accuracy of a 9-layer MLP with 512 neurons following the protocol described in \cite{regularization_is_all_you_need} (i.e., best MLP configuration) where the number of layers refers to the ``hidden'' layers of the neural network. The NAS takes this MLP as a starter and reduces the number of layers and neurons until reaching the smallest possible neural network size with minimal accuracy loss, which becomes a 3-layer MLP with 64 neurons. 

Figure \ref{plot:accuracy_pred_circuits_vs_mlp} shows the prediction accuracy of six models (i.e., XGBoost, Tiny Classifiers, non-quantized best MLP, 2-bit quantized best MLP,  non-quantized smallest MLP and 2-bit quantized smallest MLP). Across all datasets, the non-quantized best MLP model tops the performance by 83\% overall prediction accuracy whilst its 2-bit quantized version has the same performance as Tiny Classifiers. In contrast, the non-quantized smallest MLP has an overall prediction accuracy of 80\% whilst its 2-bit version stays at 75\%. In summary, the performance of Tiny Classifiers is no worse than the 2-bit quantized MLP models.

Figure \ref{plot:further_exploration} illustrates the benefit of increasing the circuit size limit from 300 gates to 400 gates of Tiny Classifiers for four datasets which present a poor classification accuracy compared to AutoGluon XGBoost. The prediction accuracy for these four datasets improves by up to 11 percentage points when moving the limit from 300 to 400 logical gates.

\subsection{ASIC Flow Results}
\label{sec:sec:evaluation_ASIC}

We design Tiny Classifiers in hardware across all datasets. For a comparison point, we also design the two ML baseline models in hardware. In addition to XGBoost (best performing ML baseline), the 2-bit quantized smallest MLP is also chosen as the second baseline ML model because it is the smallest MLP baseline (3 layers/64 neurons). As we needed to design the baseline ML models in hardware manually, we designed them only for two datasets (i.e., \emph{blood} and \emph{led}). 

These two datasets are selected based on the number of classes and the complexity of implementing XGBoost in hardware. \emph{blood} has one of the smallest numbers of classes (i.e., 2) and \emph{led} has one of the largest numbers of classes (i.e., 10). The default number of estimators (Parallel Decision Trees) for XGBoost in Python \cite{xgboost_tool} is 100 for a binary classification problem and $100 \times number\_of\_classes$ for multi-class classification. The number of estimators is strongly correlated with the achieved accuracy of the model. The main reason why \emph{blood} is selected among other 2-class datasets is because XGBoost in \emph{blood} has the smallest number of estimators with the smallest accuracy loss across all the 2-class datasets. A similar observation is made for \emph{led} that it requires a smaller number of estimators compared to \emph{yeast} (i.e., the other dataset that has also 10 classes) to achieve iso-performance.


One estimator (binary classification) for \emph{blood} and 10 estimators (one estimator for each target class) for \emph{led} are designed in hardware for XGBoost. For the development and the verification of the MLP and XGBoost designs, Bluespec System Verilog is used \cite{bluespec}, and the designs are simulated with \textit{Bluesim}.

\begin{figure*}[h!] 
    \centering
    \includegraphics[width=0.8\textwidth]{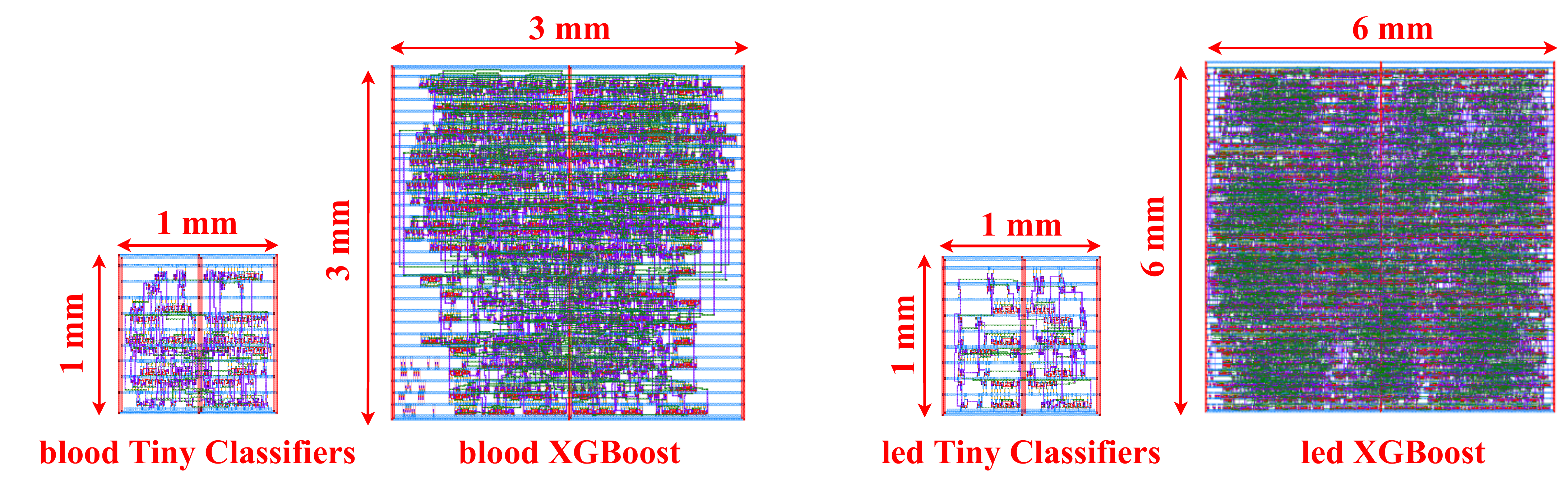}
    \caption{Flexible chip layouts of Tiny Classifiers and XGBoost implemented in PragmatIC's 0.8$\mu$m FlexIC TFT process for \emph{blood} and \emph{led} datasets.}
    \label{plot:pragmatIC_chip}
\end{figure*}

\begin{figure}[tbh!] 
    \centering
    \includegraphics[width=\linewidth]{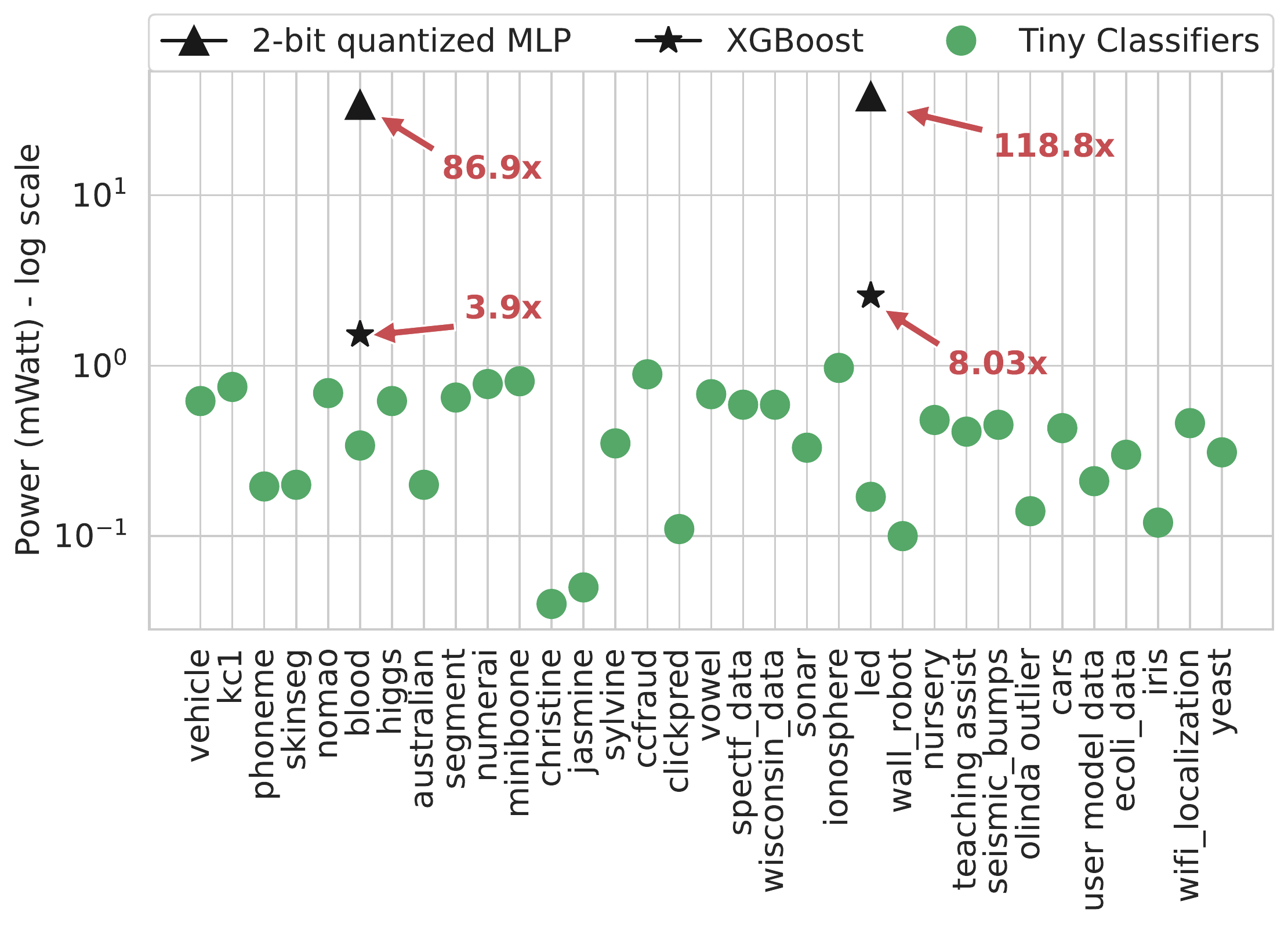}
    \caption{Power consumption of Tiny Classifier circuits across all datasets where MLP and AutoGluon XGBoost designs are shown for \emph{blood} and \emph{led} datasets.}
    \label{plot:power_asic}
\end{figure}

\subsubsection{Synthesis Results for Silicon Target:} The Verilog representation of Tiny Classifiers and the two ML baselines are synthesized using Synopsis Design Compiler targeting the open 45nm PDK \cite{freePDK45} Silicon technology. We present the synthesis power and area results for each Tiny Classifier circuit and baseline ML model as standalone hardware blocks, i.e., no interconnections to other components part of an overall ASIC design. The required data has been transferred to an input buffer, and the class predictions are stored in an output buffer inside the block. Both input and output buffers are included in the power and area calculations. The operational voltage and frequency are 1.1V and 1GHz, respectively.

\begin{figure}[h!] 
    \centering
    \includegraphics[width=\linewidth]{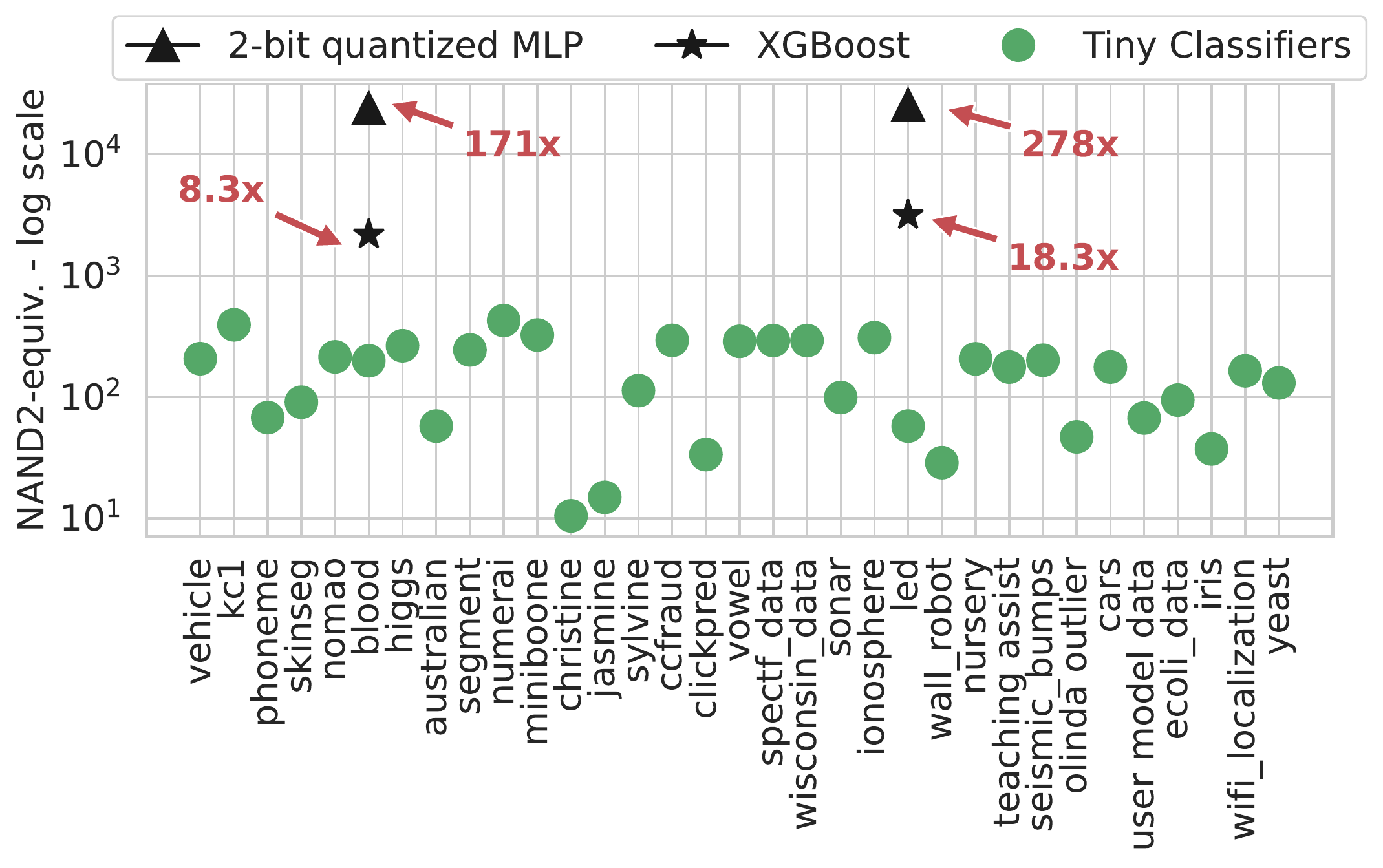}
    \caption{NAND2-equivalent gate count of Tiny Classifier circuits where MLP and AutoGluon XGBoost are shown for \emph{blood} and \emph{led} datasets.}
    \label{plot:area_asic}
\end{figure}




Figures \ref{plot:power_asic} and \ref{plot:area_asic} show the power consumption and the area in NAND2-equivalent gate count. Tiny Classifier circuits consume 0.04 - 0.97 mW, and the gate count ranges from 11-426 NAND2-equivalent gates (combinational logic plus the I/O buffers). Note that two classifier circuits have just 3 gates excluding the I/O buffers. The power consumption of MLP is 34-38 mW (86-118 times greater than that of Tiny Classifiers), and the area is $\sim$171 and $\sim$278 times larger than Tiny Classifiers for \emph{blood} and \emph{led}. The power consumption of XGBoost is $\sim$3.9 and $\sim$8 times higher than Tiny Classifiers for \emph{blood} and \emph{led} whilst the area is 8 and 18 times larger than Tiny Classifiers, respectively. 

\subsubsection{Implementation Results for Flexible Chips:} As discussed in \emph{Introduction}, Tiny Classifiers are ideal for low-cost flexible chips for smart packages. We pick XGBoost as the ML baseline for comparison because it is more efficient in terms of occupied area and power than the MLP. Both Tiny Classifiers and XGBoost designs for \emph{blood} and \emph{led} are implemented with PragmatIC's 0.8$\mu$m FlexIC metal-oxide thin-film transistor (TFT) process in PragmatIC's FlexLogIC line \cite{pragmatic}. The designs are put through the Cadence implementation flow to generate chip layouts\footnote{Tiny Classifier and XGBoost designs for \emph{blood} are sent for fabrication. They will be fabricated on a 30$\mu$m thick polyimide substrate and tested.}.

Figure \ref{plot:pragmatIC_chip} shows the flexible chip layouts of the four designs. Table \ref{tab:pragmatIC_asic_results} summarizes the power, performance and area results. Tiny Classifier for \emph{blood} is 10 times smaller and consumes about 13 times less power than XGBoost whilst it can run twice as fast as XGBoost. On the other hand, the comparative results for \emph{led} are more prominent as Tiny Classifier is about 75 times smaller \& lower power and three times faster than XGBoost. An important observation is that the area variation of Tiny Classifiers between a binary and a multi-class classification problem is negligible. Specifically, our methodology generates a smaller Tiny Classifier for \emph{led} (105 NAND2-equiv. gates) compared to \emph{blood} (150 NAND2-equiv. gates). In contrast, XGBoost implementation for \emph{led} occupies 5 times more area than \emph{blood} mainly due to the larger number of mapped estimators for multi-class classification.

\begin{table}[h!]
    \centering
    \begin{tabular}{|p{2.8cm}|cc|cc|}
    \hline
    \multicolumn{1}{|c|}{}      & \multicolumn{2}{c|}{\textbf{Tiny Classifiers}} & \multicolumn{2}{c|}{\textbf{XGBoost}}       \\ \hline
    \multicolumn{1}{|c|}{}     & \multicolumn{1}{c|}{\emph{blood}}   & \emph{led}    & \multicolumn{1}{c|}{\emph{blood}} & \emph{led}   \\ \hline
    Cell Area ($mm^2$)                & \multicolumn{1}{c|}{0.54}    & 0.37   & \multicolumn{1}{c|}{5.4}   & 27.74 \\ \hline
    Power ($mW$)                   & \multicolumn{1}{c|}{0.32}    & 0.25   & \multicolumn{1}{c|}{4.12}  & 18.6  \\ \hline
    Max. Freq. ($kHz$)               & \multicolumn{1}{c|}{350}     & 440    & \multicolumn{1}{c|}{165}   & 130   \\ \hline
    NAND2-equivalent                      & \multicolumn{1}{c|}{150}     & 105    & \multicolumn{1}{c|}{1520}  & 7780  \\ \hline
    \end{tabular}
    \caption{Tiny Classifiers and XGBoost implementation results in PragmatIC's 0.8$\mu$m FlexIC TFT process at 3V supply voltage.}
    \label{tab:pragmatIC_asic_results}
\end{table}

\subsection{FPGA-based Comparison}
\label{sec:sec:evaluation_FPGA}

We also prototype Tiny Classifiers, XGBoost and the 2-bit quantized smallest MLP for the two datasets on an FPGA platform to demonstrate the software-hardware co-design environment. Trained 2-bit quantized MLPs are synthesized on reconfigurable hardware using Xilinx FINN, the state-of-the-art tool which generates dataflow-style architectures of neural networks on FPGAs \cite{xilinx_finn}. After the initial configuration of the MLPs, we use Brevitas \cite{brevitas} to transform the neural network to a quantized trained neural network. For the Brevitas training, we use 2-bit quantized ReLU activation functions and apply batch normalization between each layer and its activation. The configuration of Brevitas follows the recommendations of Xilinx FINN \cite{xilinx_finn_with_mlp}. Then, Xilinx FINN is used to implement the trained neural network as a dataflow accelerator on FPGAs. We set the default configuration settings to build in \textit{dataflow} mode.

\begin{figure}[h!] 
    \centering
    \includegraphics[width=0.48\textwidth]{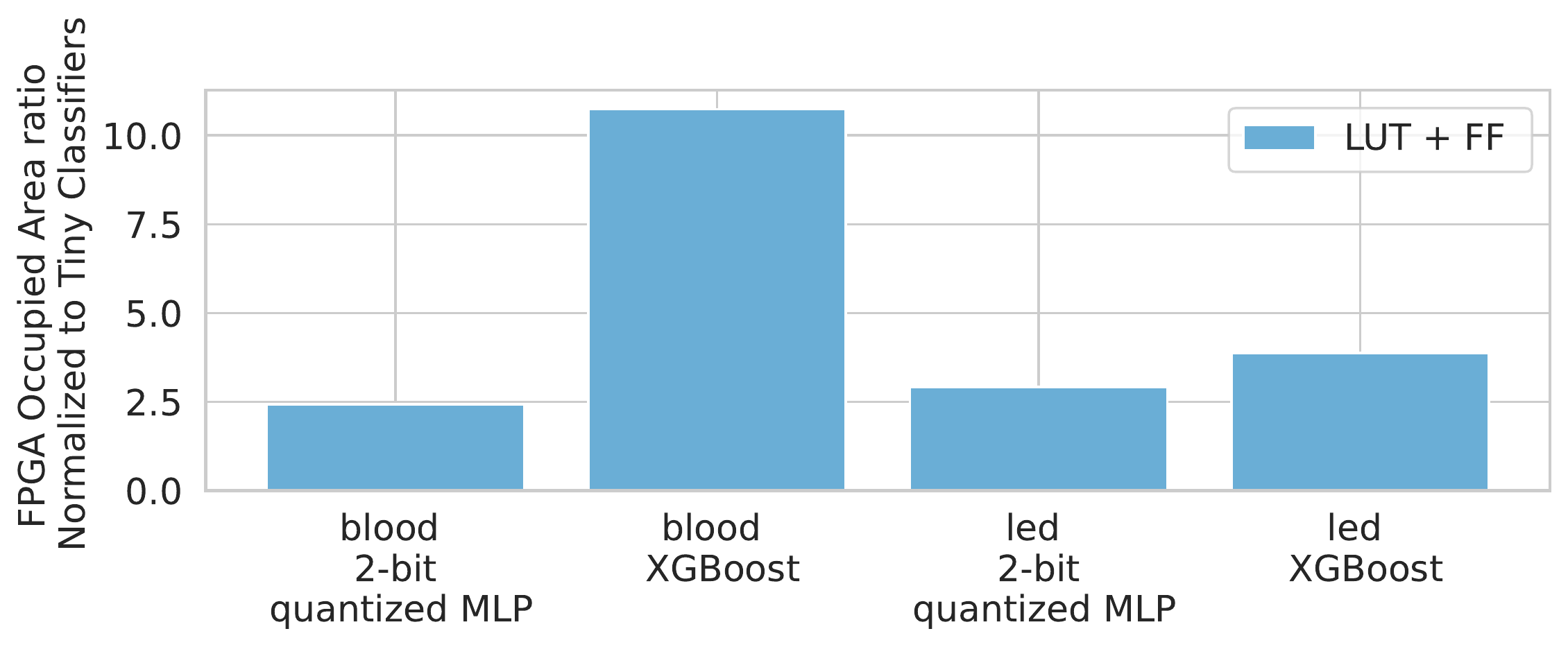}
    \caption{FPGA resource utilization (LUTs and FFs) of the 2-bit quantized Smallest MLP, XGBoost and Tiny Classifiers hardware implementation for \emph{blood} and \emph{led} datasets.}
    \label{plot:resources_fpga}
\end{figure}

Figure \ref{plot:resources_fpga} presents the FPGA resource utilization on a Xilinx Zynq Ultrascale+ MPSoC. For \emph{blood} dataset, we observe that Tiny Classifiers consume 2.43x less FPGA resources in terms of the number of look-up tables (LUTs) and flip-flops (FFs) when compared to XGBoost, and 10.7x less FPGA resources than the Smallest MLP. For \emph{led} dataset, XGBoost and the Smallest MLP are 2.92x and 3.87x larger than Tiny Classifiers, respectively.

\section{Related Work}
\label{sec:related_work}


Several methods have been proposed for supervised classification on tabular data. Two popular modern approaches are Gradient Boosted Decision Trees (GBDT), such as XGBoost \cite{chen2016xgboost} and CatBoost \cite{prokhorenkova2018catboost}, and DNNs, such as TabNet by Google \cite{tabnet} and NODE \cite{popov2019neural}. Recent work on DNNs \cite{popov2019neural, regularization_is_all_you_need} demonstrates that MLPs can be made competitive with state-of-the-art GBDT when the dimensions of the MLP architecture are suitably optimized. In addition, these optimized MLPs can also provide better accuracy than Google's TabNet \cite{tabnet}. Recall the full MLP exploration/optimization performed for our evaluation and associated accuracy results in Figure \ref{plot:accuracy_pred_circuits_vs_mlp}. This exploration ensures that the comparisons against the best MLP for accuracy, and resource utilization, against the smallest MLP, have optimized state-of-art baselines. Furthermore, using these baselines, our evaluation also facilitates the comparison of tiny classification circuits against their MLP counterparts on a well-established DNN accelerator.



Figures \ref{fig:subfig_autoML}, \ref{fig:subfig_nas} and \ref{fig:subfig_nais} highlight the main features of AutoML, NAS-based and NAIS to generate ML-based hardware accelerators.  AutoGluon is a prominent example of AutoML \cite{agtabular} \cite{agtabulardistill} as well as H2O \cite{h2o}, AutoWeka \cite{Kotthoff2019},
Auto-Sklearn \cite{Feurer2019}, MLJAR \cite{mljar_website} and Google Cloud AutoML Tables \cite{googelecloudautoML_web}. 
The current tools for AutoML do search the space for possible ML models (e.g.\ ensembles, DNNs, random forest) and these can be deployed for different inference tasks. Our experiments have used AutoGluon as a way to establish an optimized baseline for the accuracy of Neural Networks and XGBoost. However, the ML models generated by AutoML tools for tabular data do not generate RTL. That complex final step has to be done manually; see Figure \ref{fig:subfig_autoML}. For NAS tools we find a dichotomy. On one hand, we find NAS tools which can handle tabular data, but only target standard processors, GPUs, and established programmable DNN accelerators (AutoGluon, Google Cloud AutoML). On the other hand, we can find those that cannot handle tabular data but can co-design a programmable Neural Network Accelerator (NAIS approach). These rely on known ML/NN model pools and known hardware architectures  \cite{Abdelfattah_et_all_co_design} \cite{hao_et_all_co_design_2}  \cite{hao_et_all_co_design_1} \cite{Jiang_et_all_co_design}; most cases focusing on FPGAs. In the experiments, we have shown that a NAS exploration of MLPs for tabular data (as suggested by Kadra \textit{et al}.\ \cite{regularization_is_all_you_need}) produces accuracy results similar to or better than the NN produced by Amazon's AutoGluon and Google's TabNet while having the advantage of being smaller NNs.

Although our methodology aims to generate classifier circuits for tabular data, it is not in principle limited to tabular data. Work on recurrent graph-based genetic programming \cite{turner2014recurrent,atkinson2019thesis} indicates the general applicability of the evolutionary approach to other forms of data, e.g.\ time-series data. Nonetheless, making progress with Graph-Based Genetic Programming in different data domains still remains a significant research challenge in its own right.

\section{Conclusions}
\label{sec:conclusions}
This paper proposes a methodology called ``Auto Tiny Classifiers'' to automatically generate classification circuits from tabular data. We have identified a connection between Graph-Based Genetic Programming with the classification problem in ML and proposed an evolutionary approach to generate Tiny Classifier circuits composed of a small number of logic gates (i.e., < 300 gates) and capable of matching the performance of the state-of-the-art ML techniques for tabular data. 

We have evaluated the auto-generated Tiny Classifiers across 33 datasets and presented the synthesis results of Tiny Classifiers and ML baselines designed in ASIC in 45nm Silicon technology providing significant improvements in area/power. We have further implemented Tiny Classifiers and XGBoost (smallest ML baseline) as flexible chips using 0.8$\mu$m FlexIC TFT process technology. The full chip implementation results have shown that Tiny Classifiers could be clocked 2-3x faster and were 10-75x smaller and had lower power than XGBoost. We have also implemented Tiny Classifiers on an FPGA and demonstrated their area efficiency (3-11x fewer resources). 

Thus, Tiny Classifiers can be integrated as tightly-coupled functional units or co-processors or become loosely-coupled hardware accelerators. Their smaller footprint and low power consumption make them attractive for near-sensor computing and emerging smart package applications.


\section{Acknowledgements}
Konstantinos Iordanou is funded by an Arm Ltd.\ \& EPSRC iCASE PhD Scholarship. Mikel Luj\'{a}n is funded by an Arm/RAEng Research Chair award and a Royal Society Wolfson Fellowship. The research carried out by Timothy Atkinson happened while being an employee of the University of Manchester. The research is partially funded by EPSRC LAMBDA (EP/N035127/1) and EnnCore (EP/T026995/1), and UKRI NimbleAI (no.\ 10039070).

\bibliographystyle{plain}
\bibliography{main}

\end{document}